\documentclass[draft]{agujournal2019}
\usepackage{url} 
\usepackage{lineno}
\usepackage[nice]{nicefrac}
\usepackage[inline]{trackchanges} 
\usepackage{soul}
\draftfalse

\journalname{Geophysical Research Letters}

\begin{document}

%
%


\title{Depth Dependent Dynamics Explain the Equatorial Jet Difference Between Jupiter and Saturn}

%
%




\authors{Keren Duer\affil{1}, Eli Galanti\affil{1}, and Yohai Kaspi\affil{1}}


\affiliation{1}{Department of Earth and Planetary Sciences, Weizmann
Institute of Science, Rehovot, Israel}




\correspondingauthor{Keren Duer}{keren.duer@weizmann.ac.il}



\begin{keypoints}
\item Through 3D numerical simulations, we deduce that the flow depth of
gas giants significantly influences their equatorial dynamics. 
\item Equatorial flows are driven by eddy fluxes perpendicular to the axis
of rotation, and these fluxes are proportionate to the flow depth.
\item The zonal flow depth of Jupiter and Saturn, proportional to their
3:1 mass ratio, leads to the corresponding ratio in flow strength.
\end{keypoints}

%
%

%
%


\begin{abstract}
Jupiter's equatorial eastward zonal flows reach wind velocities of
$\sim100\,\,{\rm m\,s^{-1}}$, while on Saturn they are three times
as strong and extend about twice as wide in latitude, despite the
two planets being overall dynamically similar. Recent gravity measurements
obtained by the Juno and Cassini spacecraft uncovered that the depth
of zonal flows on Saturn is about three times greater than on Jupiter.
Here we show, using 3D deep convection simulations, that the atmospheric
depth is the determining factor controlling both the strength and
latitudinal extent of the equatorial zonal flows, consistent with
the measurements for both planets. We show that the atmospheric depth
is proportional to the convectively-driven eddy momentum flux, which
controls the strength of the zonal flows. These results provide a
mechanistic explanation for the observed differences in the equatorial
regions of Jupiter and Saturn, and offer new understandings about
the dynamics of gas giants beyond the Solar System.
\end{abstract}

\section*{Plain Language Summary}
In this study, we investigate the strong eastward jet around the equator
of Jupiter and Saturn. Despite the planets being similar, Saturn's
winds are stronger and cover a wider area in latitude. Recent spacecraft
measurements revealed that Saturn's winds go much deeper into its
interior than Jupiter's. Using numerical simulations, we find that
the depth of the atmosphere is crucial in determining the strength
and width of these winds. We show that the depth is related to a specific
turbulent flow, dictating the strength of the jets. These findings
explain why Jupiter and Saturn have different equatorial zonal wind
patterns and provide new insights into the behavior of giant planets
outside the Solar System.

%
%

%


%
%
%
%

\section{Introduction}

Measurements from recent decades have revealed several key similarities
between the atmospheres of the two gas planets of the Solar System,
Jupiter and Saturn. They are both dominated by east-west jet-streams
(zonal flows) surrounding the planet \cite{Tollefson2017,garcia2011},
which have been fairly stable over the past 50 years, since the Voyager
era \cite{fletcher2020}. The jets on both planets penetrate thousands
of kilometers into the interior, where the pressure exceeds $10^{5}\,{\rm \,bar}$
\cite{Kaspi2018,galanti2019}, to a depth where the Lorentz force
might become significant \cite{Liu2008,kaspi2020}. Both planets
have wide regions possessing strong turbulent activity,{ like
the famous 'storm alley' \cite{sromovsky2018} or Great White Spot
\cite{sanchez2019} on Saturn, or the Great Red Spot \cite{wong2021}
and ovals \cite{Sanchez-Lavega2001} on Jupiter}. The turbulence
at the cloud level was shown to be related to the atmospheric mean
flow, on both planets, hence the midlatitude jets are considered to
be eddy-driven \cite{Salyk2006,delgenio2012,duer2021,duer2023}.

The dynamical similarities between the planets are not surprising,
as they are both of similar size \cite{hubbard1973}, have an overall
similar composition \cite{encrenaz1999}, rotate with nearly the
same rotation rate \cite{Helled2015} and have comparable internal
heat sources \cite{guillot2022}, all of which affect the planetary
atmospheric dynamics. However, alongside the similarities, the two
planets possess notable differences. Some have comprehensive explanations
and theories, like the cloud optical depth and condensation level
\cite{atreya1999,west2004}, the temperature vertical structure \cite{mueller2008},
the penetration depth of the zonal winds, being inversely proportional
to the planetary mass \cite{kaspi2020}, and others. Nevertheless,
there remain certain unexplained disparities, with one of the most
prominent being the distinct differences observed in the equatorial
zonal flows on both planets.

\begin{figure}
\begin{centering}
\includegraphics[width=1\textwidth]{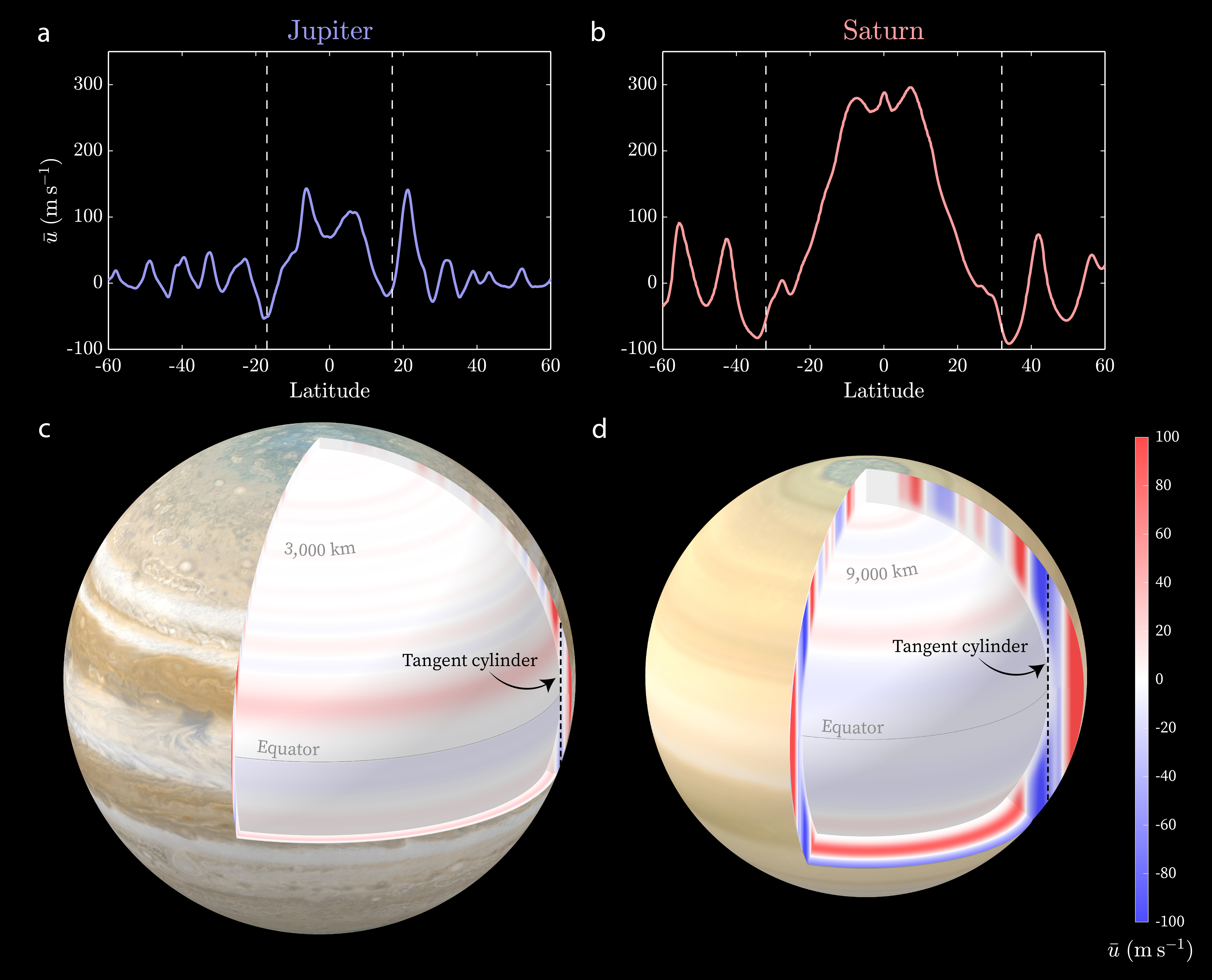}
\par\end{centering}
\caption{\label{fig:observations}The zonal winds on Jupiter and Saturn. (a)
Jupiter's and (b) Saturn's zonally-averaged zonal wind profiles \cite{Tollefson2017,garcia2011},
along with the latitude of the tangent cylinder (dashed white lines).
(c) and (d) The zonal winds of Jupiter and Saturn projected onto a
sphere with the radial decay profiles from the Juno \cite{Kaspi2018}
and Cassini \cite{galanti2019} results, respectively. Jupiter's
inner surface is at $0.955\,R_{{\rm J}}$ ($3,000\,{\rm \,km}$) and
Saturn's at $0.84\,R_{{\rm S}}$ ($9,000\,\,{\rm km}$), representing
the depth of the two atmospheres according to the gravity analysis.
The latitudinal slice is from $90^{\circ}\,{\rm N}$ to $20^{\circ}\,{\rm S}$
on Jupiter and $40^{\circ}\,{\rm S}$ on Saturn. The colorbar is shared
between panels (c) and (d) emphasizing that Saturn's zonal jets are
stronger than Jupiter's. The planets are to scale.}
\end{figure}

Recent measurements by Juno \cite{Iess2018,Kaspi2018,duer2020,kaspi2023}
and the Cassini Grand Finale \cite{iess2019,galanti2019} have provided
an estimate for the depth of the atmospheric mean flow for Jupiter
($\sim3,000\,\,{\rm km}$) and Saturn ($\sim9,000\,\,{\rm km}$),
respectively. These depths can be projected onto the $1\,{\rm bar}$
planetary surface to give the cloud-level latitude of the tangent
cylinder (Fig.~\ref{fig:observations}a,b, dashed white lines). The
tangent cylinder is an imaginary cylinder, parallel to the planet\textquoteright s
rotation axis, at a specific equatorial depth, which differentiates
the equatorial region from the midlatitudes (Fig.~\ref{fig:observations}c,d,
dashed black lines). At the equatorial region, corresponding to outside
from the tangent cylinder, both planets have a strong prograde flow
(in the direction of rotation), which is superrotating within $\sim6^{\circ}$
($\sim10^{\circ}$) latitude of the equator \cite{imamura2020} and
reaching $\sim100\,{\rm \,m\,s^{-1}}$ ($\sim300\,\,{\rm m\,s^{-1}}$)
on Jupiter (Saturn) \cite{garcia2011,Tollefson2017} (Fig.~\ref{fig:observations}a,b).
Note that the wind profile of Saturn is shown according to the latest
estimate for the rotation rate \cite{mankovich2023}. {This
estimate aligns with findings from the past two decades, indicating
Saturn's rotation period to be approximately between $10{\rm h\,}32{\rm min}$
and $10{\rm h\,}34{\rm min}$ \cite{Anderson2007,Read2009,Helled2015,mankovich2019,mankovich2023}.}
Closer to the tangent cylinder there are strong retrograde flows reaching
$\sim30\,\,{\rm m\,s^{-1}}$ on Jupiter \cite{Tollefson2017} and
$\sim90\,{\rm \,m\,s^{-1}}$ on Saturn \cite{garcia2011} (Fig.~\ref{fig:observations}a,b).
Notably, the equatorial winds on Saturn are stronger and latitudinally
wider than on Jupiter (Fig.~\ref{fig:observations}).

The extent and strength of the equatorial zonal flows on giant planets
are influenced by several factors, which have been studied using numerical
simulations and theoretical arguments \textcolor{black}{\cite{Vasavada2005,sanchez2019b}}.
The planetary rotation rate directly impacts the direction and strength
of the equatorial zonal flows, with rapidly rotating planets having
stronger prograde zonal winds \cite{Kaspi2009,gastine2013,camisassa2022}.
Heating sources and the thermal diffusivity control the convection
and the onset of convection, which in turn affects the energy transferred
to the mean zonal wind \cite{Busse1976,Christensen2001,Christensen2002,Kaspi2009}.
The depth of the convective layer has been shown to influence the
position of the tangent cylinder and hence the latitudinal extent
of equatorial dynamics \cite{busse1983,Heimpel2007,Heimpel2011,Gastine2014}.
Numerical simulations have also shown that the numerical viscosity
parameter \cite{Kaspi2009,Showman2011}, boundary conditions \cite{jones2003},
and factors such as background reference entropy state and compressibility
(and the possible presence of a stable layer) \cite<e.g.,>[]{Gastine2012,wicht2020,gastine2021,heimpel2022,wulff2022}
affect the strength of the jets in simulations, including the equatorial
region, while magnetic field effects may also play a role in the structure
of zonal winds \cite{Gastine2014,yadav2020,yadav2022}. Nonetheless,
these mechanisms cannot fully explain the differences in the equatorial
zonal flows between Jupiter and Saturn. {Notably, Jupiter exhibits
a faster rotation rate and a stronger internal heat source compared
to Saturn. While these characteristics are anticipated to enhance
cylindrically oriented convection, driving momentum through upgradient
fluxes and resulting in the generation of a stronger equatorial zonal
mean flow \cite<e.g.,>[]{Kaspi2009}, the observed flow strength
appears to be weaker on Jupiter. }Furthermore, it is likely that thermal
diffusivity, viscosity, and boundary conditions are fairly consistent
between the two planets, and as such, these factors alone cannot account
for the observed differences.

In this study, using the recent estimations of the different atmospheric
depth of Jupiter and Saturn \cite{Kaspi2018,galanti2019}, we investigate
the cause of the difference in equatorial zonal flow between Jupiter
and Saturn. Utilizing high-resolution numerical simulations, we explore
the relationship between the depth of the atmosphere and the formation
of these flows. By manipulating the domain size while keeping other
simulation parameters constant, we are able to provide an explanation
for this phenomenon. Additionally, we delve into the underlying mechanism
responsible for the formation of equatorial zonal flows on gas giants
in general.

\section{Methodology\label{sec:Numerical-simulations}}

{In this study, we investigate the equatorial zonal flows on
gas giants (including a superrotation) arising from convectively-driven,
fast-rotating conditions. Our approach involves conducting high-resolution,
hydrodynamic, anelastic simulations utilizing the Rayleigh model \cite{Featherstone2022}.}
The model solves the Magnetohydrodynamic (MHD) equations, in a rotating
frame, within spherical shells. The model was used extensively to
study Sun-like stars \cite{featherstone2016,OMara2016,karak2018,orvedahl2018},
Earth's core \cite{matsui2016,buffett2019}, subsurface oceans on
planetary moons \cite{miquel2018} and gas giants \cite{heimpel2022,duer2023}.
The model can also successfully reproduce benchmark results from other
MHD models (e.g. MagIC, Calypso) for Jupiter-like gaseous planets
\cite{Christensen2001,Heimpel2005,Jones2011,Heimpel2016}.

The hydrodynamic set of equations can be written with dimensionless
control parameters: the modified Rayleigh number $Ra^{*}=\frac{g_{o}\Delta S}{c_{p}\Omega^{2}L}$,
the Ekman number $Ek=\frac{\nu}{\Omega L^{2}}$, the Prandtl number
$Pr=\frac{\nu}{\kappa}$ and the dissipation number $Di=\frac{g_{o}L}{c_{p}T_{o}}=\eta\left({\rm e}^{\nicefrac{N_{\rho}}{n}}-1\right)$,
in addition to the polytropic index $n$ (see SI - Supporting Information),
the number of scale heights of density across the shell $N_{\rho}=\ln\left(\frac{\rho_{i}}{\rho_{o}}\right)$
and the radius ratio $\eta=\frac{r_{i}}{r_{o}}$, where the subscripts
$i$ and $o$ denote the inner and outer boundaries, respectively
\cite{Jones2011}. In the Rayleigh number, $g_{o}$ is the gravitational
acceleration at the top of the domain, $\Delta S$ is the entropy
difference across the domain, $c_{p}$ is the specific heat capacity
at constant pressure, $\Omega$ is the planetary rotation rate and
$L$ is a typical length unit. {The modified Rayleigh number
can also be formulated by incorporating the classical Rayleigh number
($Ra=\frac{g_{0}\Delta sL^{3}}{c_{p}\kappa\nu}$), such that $Ra^{*}=\frac{Ek^{2}}{Pr}Ra$.}
In the Ekman number, $\nu$ is the kinematic viscosity and in the
Prandtl number $\kappa$ is the thermal diffusivity constant, $\rho$
is the density and $r$ is radius. For the non-dimensional equations
we adopt the following units: length $\rightarrow L$ (shell depth),
time $\rightarrow\nicefrac{1}{\Omega}$, temperature $\rightarrow T_{o}$,
density $\rightarrow\rho_{o}$, and entropy $\rightarrow\Delta S$.

Using the parameters defined above we can write the non-dimensional
momentum and thermodynamic equations \cite<e.g.,>[]{Heimpel2016}:

\begin{equation}
\frac{{\ensuremath{\partial}}\mathbf{u}}{{\ensuremath{\partial}}t}+\mathbf{u}\cdot\nabla\mathbf{u}+2\hat{z}\times\mathbf{u}=Ra^{*}\frac{r_{o}^{2}}{r}S\hat{r}-\nabla\left(\frac{p'}{\bar{\rho}}\right)+\frac{Ek}{\bar{\rho}}\nabla\cdot\mathbf{D},\label{eq:momentum-non-dim}
\end{equation}

\begin{equation}
\widetilde{\rho}\widetilde{T}\left(\frac{\partial S}{\partial t}+\mathbf{u}\cdot\nabla S\right)=\frac{Ek}{Pr}\nabla\cdot\widetilde{\rho}\widetilde{T}\nabla S+\frac{EDi}{Ra^{*}}\Pi+Q_{i}.\label{eq:energy-non-dim}
\end{equation}
In the momentum equation (Eq.~\ref{eq:momentum-non-dim}), $\mathbf{u}$
is the 3D velocity vector ($u$ in the zonal direction, $v$ in the
meridional direction, and $w$ in the radial direction), $t$ is time,
$\hat{z}$ is the vertical coordinate (parallel to the axis of rotation),
$\hat{r}$ is the radial coordinate, $p$ is pressure, $\widetilde{\rho}$
is the background density, $S$ in entropy, and ${\bf D}$ is the
viscous stress tensor, such that $\mathbf{D}=2\widetilde{\rho}\left(e_{ij}-\frac{1}{3}\nabla\cdot\mathbf{u}\right)$,
and $e_{ij}=\frac{1}{2}\left(\frac{\partial u_{i}}{\partial x_{j}}+\frac{\partial u_{j}}{\partial x_{i}}\right)$.
In the thermodynamic equation (Eq.~\ref{eq:energy-non-dim}), $\widetilde{T}$
is the background temperature profile, $Q_{i}$ is a radially dependent
internal heating \cite{Jones2011}, and $\Pi$ is the viscous heating
term, where $\Pi=2\widetilde{\rho}\left(e_{ij}e_{ji}-\frac{1}{3}\left(\nabla\cdot\mathbf{u}\right)^{2}\right)$.
With the addition of the continuity equation and an equation of state
(see SI) the system of equations is complete.

Within the scope of this study, we refrain from delving into the determination
of the mechanism governing atmospheric depth, factors such as Ohmic
dissipation, stable layers, and significant density variations have
been proposed \cite<e.g.,>[]{Liu2008,Kaspi2009,Cao2017,christensen2020,wulff2022}.
Instead, our focus is directed towards examining the impact of this
depth on the resulting dynamics. Hence, we do not include magnetic
field equations and their impact on the flow fields, a unique background
entropy profile or a complicated equation of state. Our experiments
consists of a series of simulations, where we vary the domain depth
(representing the 'atmospheric' non-conductive region). In the more
shallow setups we also vary the viscosity term (Ekman number) to allow
numerical stability. As all numerical simulations solving for gas
giants, our simulations are overforced and include numerical viscosity
that is orders of magnitude larger then the molecular viscosity of
Jupiter and Saturn to allow high Rayleigh numbers \cite{Showman2011}.
Our goal is to keep the main control parameters of the model fixed,
hence we adjust the energy source, the thermal diffusivity constant
and the kinematic viscosity coefficient according to the chosen domain
depth ($L$). The chosen set of parameters is well-within the customary
values used in the benchmarks \cite{Jones2011}. For values of the
parameterization see Tabs.~S1 and S2. The simulations are calculated
until a statistical steady state is reached{ (about 1000 rotations)},
and all results are shown for either a snapshot or time-averaging
over $300$ rotations in steady state. The main three control parameters
that are kept constant between the different simulations are the modified
Rayleigh number, the Ekman number, and the fluid Prandtl number. As
we are interested in the statistical steady state solution, which
is assumed to be the current dynamical state for both Jupiter and
Saturn, the results are presented long after the onset problem, which
is influenced by the domain depth \cite{dormy2004,barik2023}. For
consistency, all of the simulations run with Jupiter's radius ($R$)
and rotation rate.

\section{Results}

\subsection{Eddies outside the tangent cylinder\label{subsec:Eddies-outside-the}}

\begin{figure}
\begin{centering}
\includegraphics[width=1\textwidth]{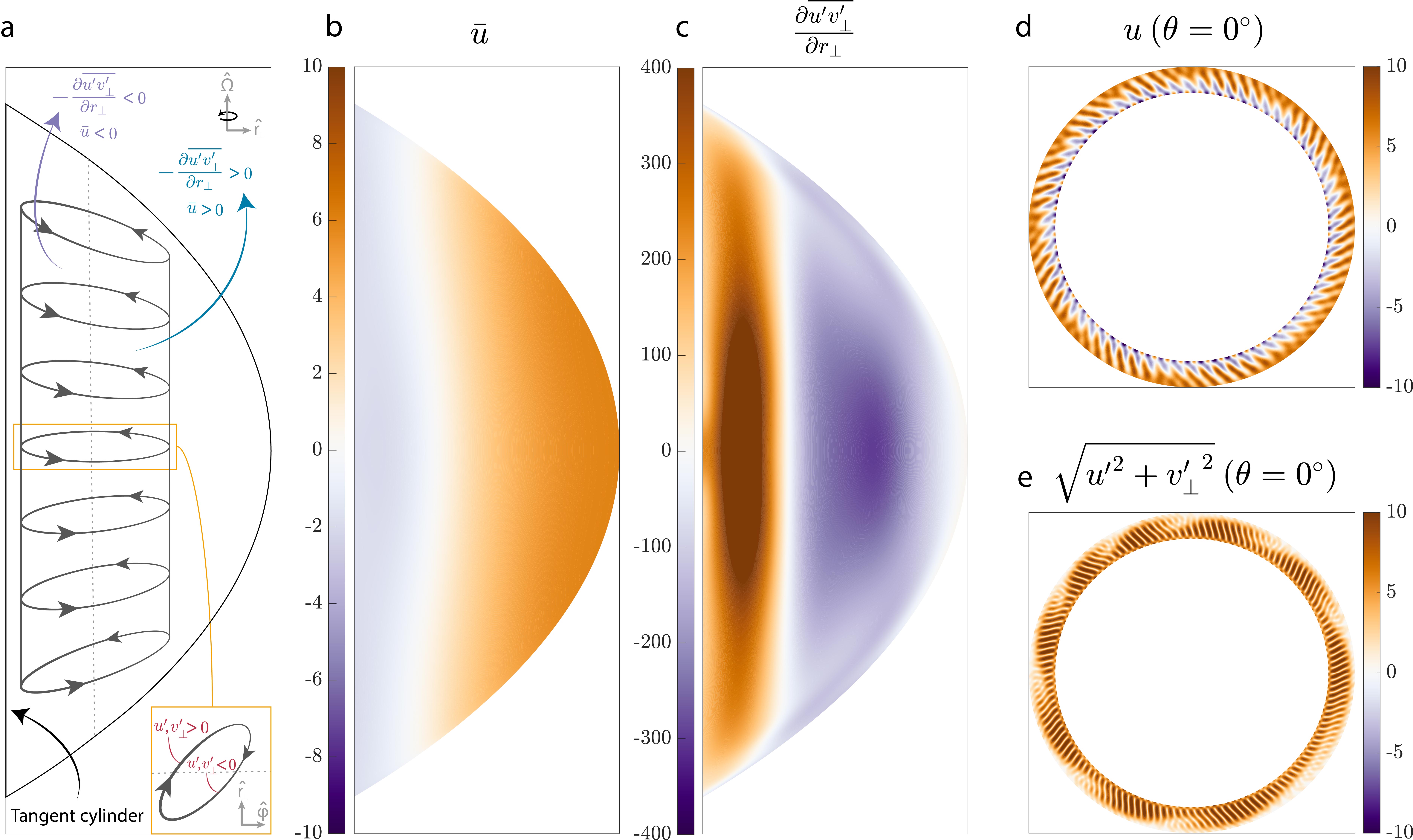}
\par\end{centering}
\caption{\label{fig:eddies}Snapshot of the dynamics in an ideal simulation
with $r_{\min}=0.84R$. (a) An illustration of a tilted convection
column (the tilt is in the positive zonal direction, i.e., into the
page, see inset), pushing positive momentum in a direction perpendicular
to the axis of rotation. The dashed black line is aligned with the
zero values in panels b and c. Due to the tilt, the perpendicular
momentum flux is always positive $\left(u^{\prime}v_{\perp}^{\prime}>0\right)$
inside the columns (see inset). (b) The zonally-averaged zonal wind
(${\rm m\,s^{-1}}$) and (c) the eddy momentum flux convergence (Eq.~S7,
${\rm m\,s^{-2}}$) that is perpendicular to the axis of rotation,
outside the tangent cylinder (the abscissa range is $[0.84-1]\,\,\nicefrac{r}{R}$).
It is evident that the zonal wind is proportional to the eddy momentum
flux convergence. (d) and (e) A snapshot demonstrating the zonal wind
and the eddy velocity at the equatorial plane (${\rm m\,s^{-1}}$),
respectively. An equivalent figure for a more turbulent simulation
with the same domain depth is provided in the SI (Fig.~S1).}
\end{figure}

{Convection in fast-rotating spherical bodies tends to align
in cylinders parallel to the rotation axis. These cylinders are tilted
in the direction of rotation and, hence, transfer positive momentum
outward, driving the equatorial zonal flows} \cite<e.g.,>[]{Busse1976,Busse1982b,Busse2002,Zhang1987}{.}
To investigate this mechanism, we focus on the convergence of eddy
fluxes oriented perpendicular to the axis of rotation within the entire
region outside the tangent cylinder (Fig.~\ref{fig:eddies}). Examining
the zonally-averaged zonal wind (Fig.~\ref{fig:eddies}b) along with
the convergence of the zonally-averaged perpendicular eddy momentum
flux (Eq.~S7, Fig.~\ref{fig:eddies}c) in an idealized simulation
($r_{\min}=0.84R$, $Ek=5\times10^{-4}$, $Pr=2$ and $Ra^{*}=0.0132$),
reveals that the zonal wind is being driven by these fluxes, which
are aligned in cylinders throughout the entire domain (except at the
outer boundary near the equator and near the tangent cylinder). Thus,
the leading momentum source in the zonal momentum equation can be
described by
\begin{equation}
\frac{\partial\bar{u}}{\partial t}\propto-\frac{\partial\overline{u^{\prime}v_{\perp}^{\prime}}}{\partial r_{\perp}},\label{eq:3}
\end{equation}
where $v_{\perp}^{\prime}=w^{\prime}\cos\theta+v^{\prime}\sin\theta$
and $\theta$ is the latitude (see SI for the full equation in an
anelastic form in spherical coordinates, Eq.~S6, S7). The bar represents
a zonal average and the prime denotes deviations from that average.
The zonal momentum equation describing such dynamics includes additional
forces, like the Coriolis force, the mean eddy fluxes and the viscosity
term, none of which dictate the direction of the zonal wind (the Coriolis
force cancels itself in the equatorial region, the mean eddy fluxes
are small due to the flow being geostrophic and the viscosity terms
acts against the reminder, \cite{Kaspi2009,duer2023}, see SI).

Examining the instantaneous zonal wind reveals the structure and number
of the convection columns throughout the simulation (Fig.~\ref{fig:eddies}d).
As this simulation is calculated with a relative high Ekman number
($Ek=5\times10^{-4}$), the resulting columns are near ideal and spread
from the domain's inner boundary to the domain's outer boundary (this
can be seen in the eddy velocity field, Fig.~\ref{fig:eddies}e).
A more turbulent simulation for a domain with the same depth, in which
the eddies are stronger but less organized, and they do not span the
entire depth of the domain, is shown in Fig.~S1. Additionally, the
zonally-averaged convergence term changes sign at multiple depths
(Fig.~S1c). Yet, averaging the simulation over time shows that the
momentum source keeps its sign and strength near the tangent cylinder
(inner boundary) and near the equator (outer boundary), as these regions
allow only tilted columns that diverge and converge, respectively.
The middle region averages to small values, as convection columns
are being built and destroyed while the simulation continues (Fig.~S1c).

The columnar structure of the flow is illustrated in Fig.~\ref{fig:eddies}a,
showing the transition between the divergence and convergence of the
eddy momentum fluxes together with the zonal wind direction dictated
by it. As the divergence of the eddies occurs outside the tangent
cylinder, the retrograde flanks of the zonal flow must also be positioned
outside the tangent cylinder. This implies that the atmospheric depth
needs to match the depth of the retrograde flow adjacent to the prograde
equatorial jet in superrotating giants.

\subsection{The extent of the equatorial zonal flows \label{subsec:The-extent-of}}

\begin{figure}
\begin{centering}
\includegraphics[width=1\textwidth]{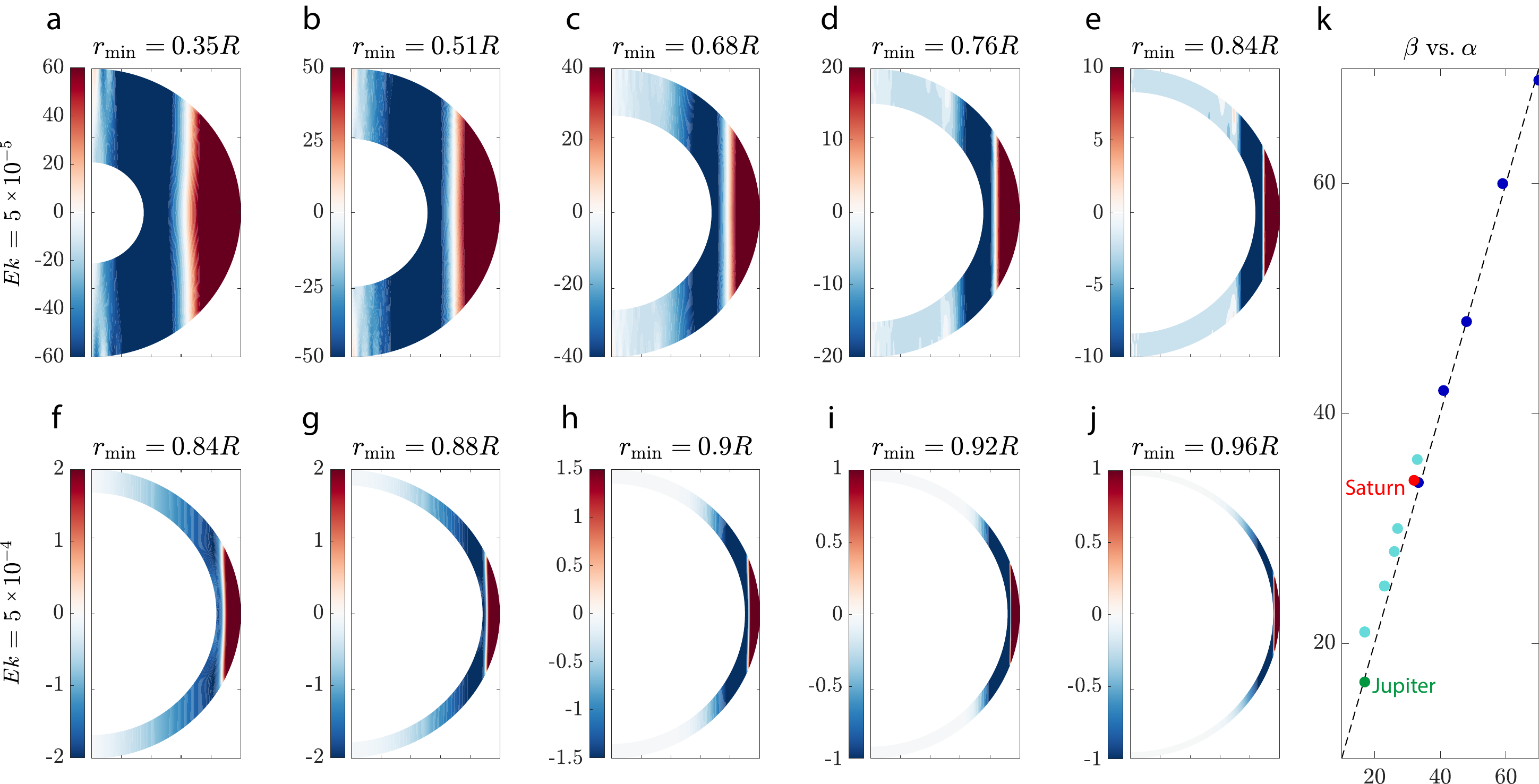}
\par\end{centering}
\caption{\label{fig:10depth} (a) - (e) Zonally averaged zonal wind (colors,
${\rm m\,s^{-1}}$) of 5 simulations dominated by a prograde equatorial
jet and an adjacent retrograde jet, each extend to a different depth
($r_{{\rm min}}$). All the simulations are calculated with identical
control parameters: $Ek=5\times10^{-5}$, $Pr=2$ and $Ra^{*}=0.0132$.
(f-j) Shallower 5 simulations with equal $Pr$ and $Ra^{*}$, but
with $Ek=5\times10^{-4}$. (k) $\beta$ (ordinate) and $\alpha$ (abscissa),
representing the latitude where the retrograde jet peaks in each simulation
and the projection of $r_{{\rm min}}$ (the tangent cylinder) on the
outer boundary, respectively. The ten simulations are presented (top
- blue, bottom - cyan) along with the values for $\beta$ and $\alpha$
of Jupiter (green) and Saturn (red). The $r^{2}$ value of panel (k)
is $0.99$. The dashed line represents the $1:1$ ratio.}
\end{figure}

Next, we examine how the depth of the atmospheric flow affects the
extent of the equatorial zonal winds. The steady-state, zonally-averaged,
zonal wind field is compared between ten different simulations, calculated
with different domain depths (Fig.~\ref{fig:10depth}). All the simulations
in the top row (Fig.~\ref{fig:10depth}a-e) are calculated with identical
control parameters: $Ek=5\times10^{-5}$, $Pr=2$ and $Ra^{*}=0.0132$
and a scaled reference state (see SI), while adjusting the other system
constants (see SI for details). For domains smaller than $r_{{\rm min}}=0.85R$
the Ekman number must be larger to constrain the eddy-related phenomena
inside the tangent cylinder (poleward from the tangent cylinder),
allowing to reach a dynamical steady state. Hence, the bottom five
simulations (Fig.~\ref{fig:10depth}f-j) are calculated with a larger
Ekman number, and their control parameters are: $Ek=5\times10^{-4}$,
$Pr=2$ and $Ra^{*}=0.0132$. For comparison, the domain of $r_{{\rm min}}=0.84R$
is presented both with both Ekman numbers.

The latitudinal extent of the equatorial zonal wind field is compared
between the different simulations, and also to Jupiter and Saturn
(Fig.~\ref{fig:10depth}k). It is defined according to the projection
of atmospheric region on the outer radius (the tangent cylinder),
represented by the latitude $\alpha$. In the simulations, it can
simply be set according to the shell's depth ($L=R-r_{{\rm min}}$)
and the chosen planetary radius ($R$), such that $\alpha=\arccos\left(1-\frac{L}{R}\right)$.
For Jupiter and Saturn, we set $L$ according to the decay profiles
provided from the gravity analysis, $3,000\,\,{\rm km}$ for Jupiter
and $9,000\,\,{\rm km}$ for Saturn. This gives values of $\alpha_{{\rm Jup}}=17^{\circ}$
and $\alpha_{{\rm Sat}}=32^{\circ}$ (dashed white lines, Fig.~\ref{fig:observations}a,b).
The latitude $\alpha$ separates the prograde equatorial zonal wind
and partly the retrograde flow engulfing it from the alternating jets
at midlatitudes (Fig.~\ref{fig:observations}c,d). Specifically,
it nearly coincides with the maximal retrograde flow, on both planets
(Fig.~\ref{fig:observations}a,b). For comparison with the simulations,
we define the latitude $\beta$, which is the latitude where the retrograde
jet adjacent to the equatorial prograde flow peaks at the outer boundary.
This determination represents well the extent of the equatorial dynamics,
since tilted convection columns are the momentum source of the equatorial
zonal jets as detailed in the previous section. For asymmetric simulations
(or observations), an averaged value was taken between the hemispheres.
This gives values of $\beta_{{\rm Jup}}=16.6{}^{\circ}$ and $\beta_{{\rm Sat}}=34.1{}^{\circ}$.
For both planets the values of $\alpha$ and $\beta$ are nearly equal.
Presenting them together with the ten simulations of Fig.~\ref{fig:10depth},
it is apparent that all ten simulations also exhibit near equal values
of $\alpha$ and $\beta$, with a $R^{2}$ value of $0.99$.

{The robustness of our findings was assessed by varying the
main parameters to ensure the general applicability of the results.
Particularly, we conducted experiments by scaling the values of $Pr$
and $Ra^{*}$ across four different domain depths, and reexamined
the correlation depicted in Fig.~\ref{fig:10depth}k for the additional
16 simulations. The results proved to be robust under these variations,
as illustrated in Fig.~S2 where the values of $\alpha$ and $\beta$
maintain the $1:1$ ratio. This emphasizes the consistent connection
between domain depth and the ensuing equatorial dynamics.}

Fig.~\ref{fig:10depth} reveals the dominance of the domain depth
in controlling the latitudinal extent of the equatorial zonal flows
and the adjacent retrograde jets. It also shows, that the two retrograde
jets adjacent to the prograde jet have the same origin and are linked
outside the tangent cylinder. This relationship has been presented
in previous studies \cite<e.g.,>[]{busse1983,Christensen2001,Heimpel2007,Gastine2014},
however, the comparison with observations for Jupiter and Saturn,
which only became recently available, constrain the position of the
tangent cylinder and allow fixating on the relative zonal winds region
that participates in the equatorial dynamics. Note that the latitude
of the tangent cylinder ($\alpha$) of the deepest simulation reaches
nearly $70^{{\ensuremath{\circ}}}$ latitude, leaving no place
for midlatitude dynamics. This could explain the absence of midlatitude
jets, for example, in deep, convection-driven, simulations \cite<e.g.,>[]{Kaspi2009}.
The columnar structure of the deep flow, characterized by Taylor columns,
is uninterrupted in the equatorial region and the adjacent retrograde
jet up to its maximal absolute value. This means that the equatorial
jet and its adjacent jets are close to north-south symmetric and continue
throughout the planet's interior, while the density varies at different
depths. Electrical conductivity, of course, changes with depth as
well \cite{French2012,Liu2008}, and might be directly linked to
what sets the depth participating in the equatorial dynamics \cite{kaspi2020}.
However, even without the magnetic field constraints, the agreement
between the simulations and recent estimations for Jupiter and Saturn
implies that the latitudinal extent of the equatorial zonal flows
is dictated directly by the atmospheric depth. 

\subsection{The strength of the zonal flows and the eddies driving it \label{subsec:Strength}}

\begin{figure}
\begin{centering}
\includegraphics[width=0.8\textwidth]{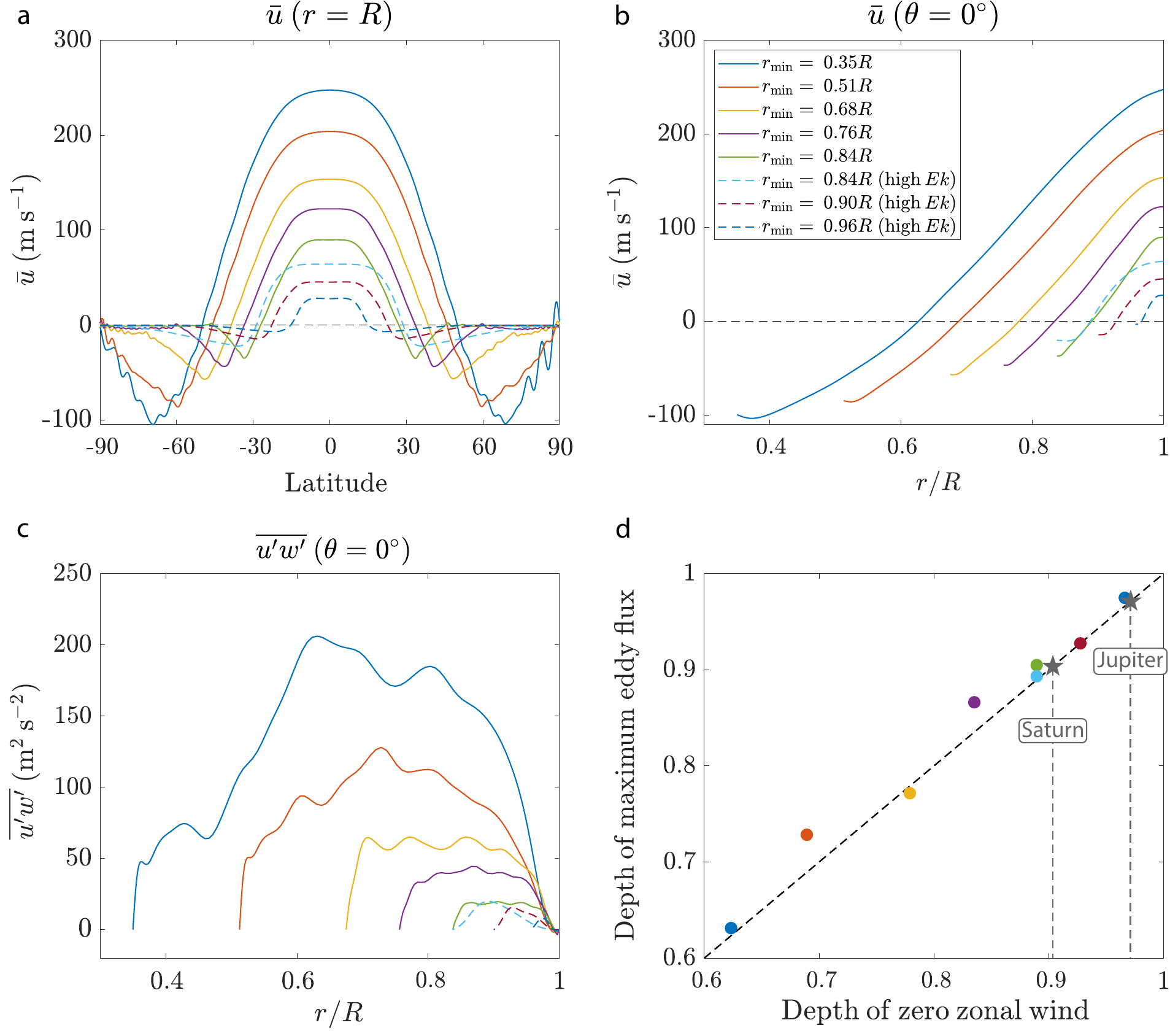}
\par\end{centering}
\caption{\label{fig:8_simulations}(a) The zonally-averaged zonal wind as a
function of latitude at the outer radius ($r=R$) of the five top
simulations (Fig.~\ref{fig:10depth}a-e, solid lines) and three additional
simulations with higher Ekman number (Fig.~\ref{fig:10depth}f,h,j,
dashed lines). The three simulations with high Ekman number are scaled
(the velocities are multiplied by $6$) such that the eight simulations
are presented in the same plot. All eight simulations have identical
Rayleigh and Prandtl numbers. (b) The zonally-averaged zonal wind
as a function of normalized radius at the equator ($\theta=0^{\circ}$)
of the same eight simulations. (c) The vertical eddy momentum flux
$\left(\overline{u^{\prime}w^{\prime}}\right)$ with normalized radius
at the equator of the eight simulations. (d) Comparison between the
depth where the vertical eddy momentum flux is maximal (ordinate)
and the depth where the zonal wind crosses the zero velocity (abscissa).
The dashed line represent the $1:1$ ratio. The $r^{2}$ value of
the eight simulations is $0.99$. Also shown are the abscissa values
for Jupiter \cite{Kaspi2018} and Saturn \cite{galanti2019} (gray),
according to the flow decay profiles obtained by the gravity measurements,
projected on the $1:1$ ratio. All simulations are at a statistical
steady state and the terms shown are time-averaged. The legend is
shared between the panels.}
\end{figure}

Lastly, we examine the strength of the equatorial zonal winds. It
is evident that in shallower domains the zonal winds weaken (Fig.~\ref{fig:10depth}).
As the main three control parameters in the simulations are kept constant
(besides the Ekman number, which is increased for simulations shallower
than $r_{{\rm min}}=0.84R$), the zonal wind velocity must be dependent
on other factors that are different between the simulations. We examine
eight simulations together on the same panels (Fig.~\ref{fig:8_simulations}),
five deep simulations (Fig.~\ref{fig:10depth}a-e) and three shallow
simulations with higher Ekman number (Fig.~\ref{fig:10depth}f,h,j).
As before, the simulations with $r_{{\rm min}}=0.84R$, are presented
with both Ekman values. The zonal winds of the high Ekman simulations
are significantly weaker than those of the deep simulations, reaching
wind velocities of a few meters per second at the equator. This is
an expected outcome as the viscosity was increased by an order of
magnitude, the zonal winds should be reduced by approximately the
same ratio (the viscosity acts against the zonal wind acceleration
term, see Eq.~\ref{eq:momentum-non-dim} and SI). To present all
eight simulations together, the three simulations with the high Ekman
value are scaled (the velocities are multiplied by a factor of $6$;
dashed lines, Fig.~\ref{fig:8_simulations}).

The wind velocity appears to be linearly proportional to the simulation
depth at all latitudes outside the tangent cylinder (Fig.~\ref{fig:8_simulations}a)
and at all simulation depths (Fig.~\ref{fig:8_simulations}b). This
suggests that the strength of the zonal winds must be directly influenced
by the size of the domain, allowing more momentum to be transferred
to the acceleration term of the zonal wind (Eq.~\ref{eq:momentum-non-dim}).
In fast-rotating, convection-driven simulations, such as the simulations
presented here, the mechanism driving momentum towards the equatorial
zonal wind is the tilted convection columns that originate outside
the tangent cylinder, as shown in Fig.~\ref{fig:eddies}. At the
equatorial plane, this can be represented by the radial eddy momentum
flux ($\overline{u^{\prime}w^{\prime}}$, Fig.~\ref{fig:8_simulations}c).

In idealized simulations, the columns are spread almost evenly through
the entire plane between the boundaries (e.g., Fig.~\ref{fig:eddies}d,e).
The tilt of the columns is in the prograde direction, hence, the eddy
momentum flux is always positive (see illustration, Fig.~\ref{fig:eddies}a).
In more turbulent simulations, the columns are noisy and continue
to break and reappear (Fig.~S1d,e). Examining the eddy momentum flux
at the equatorial plane reveals that indeed this term is always positive,
and while in the more turbulent simulations it is more noisy (even
when looking at time-averaged values), the convection columns are
positioned close to the tangent cylinder and are dominant between
the boundaries (Fig.~\ref{fig:8_simulations}c). The eddy momentum
fluxes are also proportional to the depth of the simulation. Deeper
simulations allow stronger eddies to evolve, thus, giving more momentum
to the zonal wind, resulting in higher velocities. This explains why
Saturn's equatorial winds (both prograde and retrograde) are stronger
than Jupiter's, as the tangent cylinder is positioned three times
deeper (Fig.~\ref{fig:observations}). However, due to lack of measurements
of eddy fluxes beneath the cloud level, comparison to the real atmospheres
is unavailable.

The transition in the zonal wind direction, from retrograde to prograde
(Fig.~\ref{fig:8_simulations}b), should be positioned where the
eddy momentum flux term changes its tendency, as the full term that
appears in the momentum equation is the convergence of the eddy fluxes
(Eq.~\ref{eq:3}). To examine this, we calculate the correlation
between the location where the zonal wind changes its direction ($\bar{u}=0$)
and the location where the eddy momentum flux is maximal (equivalent
to zero convergence). These values in all the simulations are very
close to the $1:1$ ratio, with $r^{2}$ value of $0.99$ (Fig.~\ref{fig:8_simulations}d),
suggesting that the zonal wind field strength is directly related
to the eddy momentum fluxes at the equatorial plane. While comparison
with Jupiter and Saturn is unavailable, given that we do not have
measurements of eddy momentum fluxes at depth, they are projected
on the $1:1$ ratio according to the abscissa (Fig.~\ref{fig:8_simulations}d),
which can be found according to the flow decay profiles obtained by
the gravity measurements by Juno \cite{Kaspi2018} and Cassini \cite{galanti2019}.
This relationship implies on the deep structure of eddies in the Jovian
and Saturnian atmospheres.

\subsection*{Conclusions\label{sec:Conclusions}}

In this study, we show that the penetration depth of the atmospheric
dynamics directly influences the extent and strength of the cloud-level
equatorial zonal flows on Jupiter and Saturn. The latitudinal extent
of the equatorial zonal flows, defined here as the latitude of the
maximum retrograde flows engulfing the prograde equatorial zonal wind,
is dictated by the depth of the atmospheric dynamics and is compatible
with the cloud-level wind structure and the zonal jets' depth given
by the gravity measurements for Jupiter and Saturn. The depth of the
atmospheres on these planets is inversely proportional to the mass
of the planets. Jupiter's mass is approximately three times heavier
than Saturn's, leading to denser gas closer to the outer radius of
the planet \cite{Guillot1999b}. The planets' mass affects the depth
at which the gas becomes ionized and the potential depth for a stable
layer to exist, hence, influencing the extent of the planetary atmospheric
depth regardless of what mechanism might dissipate the zonal jets
\cite{Liu2008,Liu2010,christensen2020,gastine2021,wulff2022}.

The strength of the equatorial zonal flows might depend on parameters
such as rotation rate, viscosity parameter, internal heating, and
boundary conditions. Yet, with a representative set of control parameters
we show that the depth of the tangent cylinder is linearly proportional
to the strength of the eddy momentum flux, and hence to the strength
of the zonal flow. This suggests that the difference between the equatorial
zonal flows of Jupiter and Saturn could be a direct result of their
atmospheric depth, mainly due to the similarities in the other parameters
that might affect the zonal flow strength. The relation between Jupiter's
and Saturn's equatorial zonal wind velocity is $1:3$, which is the
same ratio as their atmospheric depth. {Although the 1:3 ratio
for Jupiter's zonal wind strength compared to Saturn's might subtly
vary depending on rotation rate and specific measurements \cite{sanchez2023},
all wind profiles consistently support the notion that Saturn's winds
are demonstrably stronger and deeper than Jupiter's.} These results
cannot apply to Uranus and Neptune, as they are not superrotating
and possess much shallower atmospheres \cite{Kaspi2013c,soyuer2020},
which might not even hold convection columns. 

We also show that the zonal flows outside the tangent cylinder are
aligned in cylinders, along with the eddy momentum fluxes that drive
them. Deeper atmospheres allow stronger eddies in the direction perpendicular
to the axis of rotation, transferring more momentum to the zonal flows.
This mechanism is restricted to the region outside the tangent cylinder,
and cannot explain the midlatitudes and polar dynamics. The compatibility
of these modeling results with the winds on Jupiter and Saturn suggests
that tilted Taylor convection columns control the dynamics in gas
giant atmospheres outside the tangent cylinder, as suggested decades
ago \cite{Busse1976}. Lastly, based on the results presented here,
future estimation of the cloud-level zonal winds of superrotating
exoplanets may allow to constrain their radial atmospheric depth,
and hence, better understand their interior density structure and
evolution.

\section*{Open Research Section}
The data
presented in the introduction is publicly available in 
 \citeA{Tollefson2017}, \citeA{garcia2011}, \citeA{Kaspi2018} and \citeA{galanti2019}. The numerical results presented in this study were obtained by solving the hydrodynamic equations via the Rayleigh convection model, see \citeA{Featherstone2022}, with certain parameters and reference state (see main text and SI).

\section*{Conflict of Interest Statement}
The authors have no conflicts of interest to declare.

\acknowledgments

The cloud maps used in Fig.~\ref{fig:observations} are available
in https://www.solarsystemscope.com/textures/ and in https://www.planetary.org/space-images/merged-cassini-and-juno
for Saturn and Jupiter, respectively. We acknowledge the support of
the Israeli Space Agency, the Israeli Science Foundation (Grant 3731/21),
and the Helen Kimmel Center for Planetary Science at the Weizmann
Institute of Science.

%
\bibliography{Kerensbib}

%


%
%
%
%
%

\end{document}